\documentclass[aps,twocolumn,floatfix,bibnotes,altaffilletter]{revtex4}

\usepackage{dcolumn}
\usepackage{color}
\usepackage{amsmath}
\usepackage[dvips]{graphicx}

\newcommand\pictc[5]{\begin{figure}
                       \centerline{
                       \includegraphics[width=#1\columnwidth]{#3}}
                   \protect\caption{\protect\label{fig:#4} #5}
                    \end{figure}            }
\newcommand\pict[4][1.]{\pictc{#1}{!tb}{#2}{#3}{#4}}
\newcommand\rpict[1]{\ref{fig:#1}}

\newcommand{\be}{\begin{equation}}
\newcommand{\ee}{\end{equation}}
\newcommand{\bea}{\begin{eqnarray}}
\newcommand{\eea}{\end{eqnarray}}

\newcounter{Fig}

\begin{document}

\begin{sloppy}

\title{Beam oscillations and curling in chirped periodic structures with metamaterials}

\author{Arthur R. Davoyan, Andrey A. Sukhorukov, Ilya V. Shadrivov, and Yuri S. Kivshar}

\affiliation{Nonlinear Physics Center, Research School of Physical
Sciences and Engineering, Australian National University, Canberra
ACT 0200, Australia}


\begin{abstract}
We study the propagation of electromagnetic waves in one-dimensional chirped periodic structures composed of alternating
layers of negative-index (or left-handed) metamaterial and conventional dielectric, under the condition of the zero average
refractive index. We consider the case when the periodic structure has a linear chirp, and the chirp is introduced by varying
linearly the thickness of the layers across the structure. We apply an asymptotic analytical method for the analysis of the
Bloch oscillations and find that, in a sharp contrast to ordinary periodic dielectric structures, the energy flow in
multi-layered stacks with metamaterials may have the opposite direction at the band edges, thus providing novel
possibilities for the beam steering in the transmission band.

\end{abstract}

\maketitle

\section{Introduction}

Materials with negative index of refraction, also known as left-handed metamaterials, are attracting a great scientific interest nowadays. This novel type of materials was predicted theoretically back to 1967~\cite{Veselago:1967-2854:SPSS} but only 30 years later this theoretical curiosity was followed by the first experimental demonstrations based on the fabrication of composite structures consisting of split-ring resonators and metallic wires~\cite{Shelby,Smith:2004-788:SCI}. The basic properties of the left-handed metamaterials predicted theoretically have been confirmed by experiment, and this led to a rapid progress in this field and many other interesting discoveries. In particular, many new physical phenomena based on the concept of metamaterials and negative index of refraction have been predicted, including the fundamental concept of perfect lens~\cite{Pendry:2000-3966:PRL} and, more recently, electromagnetic cloaking~\cite{cloak_1,cloak_2, cloak_3}.

One of the interesting directions in the analysis of the unusual properties of this novel composite metamaterials is the study of periodic structures (or photonic crystals) composed of metamaterials. Photonic crystals, being widely used for both light control and beam manipulation, demonstrate a variety of novel physical effects, and they may offer many new possibilities being combined with metamaterials. In particular, a series of recent studies revealed the existence of a novel zero-index bangaps~\cite{Shadrivov:2003-3820:APL,Shadrivov:2005-193903:PRL,Wu:2003-235103:PRB,Li:2003-083901:PRL} associated with one-dimensional photonic crystals composed of alternating layers of negative-index and conventional dielectrics for which the averaged refractive index vanishes. Such novel periodic structures demonstrate many intriguing properties, including substantial suppression of the Anderson localization and long-wavelength resonances~\cite{anderson}.

In this paper, we study the propagation of electromagnetic weaves in one-dimensional periodic structures composed of alternating layers of left-handed metamaterial and conventional dielectric. It is well established~\cite{Bloch_oscill,Wilkinson:2002-056616:PRE} that introducing a chirp into a one-dimensional periodic structure leads to the generation of optical Bloch oscillations which can be also predicted in the presence of metamaterials~\cite{our_oe}. In addition to our numerical analysis of the Bloch oscillations studied earlier~\cite{our_oe}, in this paper we demonstrate a number of novel, unique features of the beam propagation in this kind of chirped periodic composite structures. In particular, in contrast to Ref.~\cite{our_oe}, here we consider the chirped structures with a large number of layers and slowly varying period. Applying the approximation based on the geometric optics, we reveal a variety of novel effects that could be observed in such structures, including the reversal of the energy flow at the band edges and the fascinating effect of the beam curling. This novel effect of the beam curling can be useful for the beam steering in the transmission band. We also study the eigenvalue problem and the corresponding eigenmodes and determine both the field structure and Poynting vector distribution associated with the beam propagation.

The paper is organized as follows. In Sec.~\ref{sec_structure} we introduce our structure as a one-dimensional stack of layers composed of alternating layers of metamaterial and dielectrics. We discuss the way of presenting eigenmodes of the structure with a slowly varying period and discuss the initial conditions. In Sec.~\ref{sec_periodic} we implement the Bloch theorem and derive the dispersion relation for the Bloch wave vector. Then, we analyze the bandgap diagram for this structure and discuss the conditions for the observation of a novel type of the beam dynamics associated with the Bloch oscillations.  Section~\ref{sec_geometry} is devoted to the geometric optics approximation. Here we demonstrate a possibility to apply the geometric optics for the analysis of layered structures with slowly changing parameters. We derive the equations of motion for the paraxial beams, and also study the period of oscillations pointing out novel properties of the beam trajectories connected with a change of the sign of the spatial group velocity in the structure due to the presence of a metamaterial. Also, we analyze different regimes of the beam propagation and compare them with the well-known case corresponding to the similar structures composed of conventional dielectric layers. In Sec.~\ref{sec_field}, we discuss the eigenvalue problem and point out the difficulties and failure of the commonly used numerical techniques. In addition, we describe an alternative method to find the eigenmodes of the structures with a large number of periods. Based on this method, we study the field distribution in the structure and determine the direction of the Poynting vector in the beam. We observe a close correspondence with predictions of the geometric optics. Finally, Sec.~VI concludes the paper.

\section{Structure geometry and eigenmode formalism}
\label{sec_structure}

We study a one-dimensional chirped periodic structure shown schematically in Fig.~\rpict{fig_stacks}, where the slabs of a negative-index metamaterial with the width $d_lm$ are separated by the layers of the conventional dielectric with the width $d_rm$.
We describe the variation of the refractive index in the m-th pair of layers as follows,
\be \label{eq_refractive_index}
n(z)=\left\{
        n_r = \sqrt{\varepsilon_r \mu_r} \; \; \; \; \; \; \;
                z\in(z_m,z_m+d_{rm}) \atop
        n_l = -\sqrt{\varepsilon_l \mu_l} \; \; \; \; \;
        z\in(z_m+d_{rm},z_m+\Lambda_m)\right.
\ee
where $n_l$ and $n_r$ are the refractive indices of metamaterial and
dielectric layers, respectively, $\Lambda_m$ is the width of the m-th unit
cell. In contrast to our previous studies~\cite{our_oe}, here we analyze the structures with a large number of unit cells and slowly varying period.

We consider TE-polarized waves with the electric field described by one component ${\bf E}=(E_x,0,0)$, and the waves propagating in the $(y,z)$ plane. In this case the problem is described by the Helmholtz equation,
\be \label{Helmholtz}
\Delta_2 E_x(y,z)+ n(z)^2 E_x(y,z) -\frac{1}{\mu} \frac{d\mu(z)}{dz}\frac{\partial E_x(y,z)}{\partial z}=0,
\ee
where $\Delta_2$ is the two-dimensional Laplacian. The field is assumed to be
monochromatic with the frequency $\omega$, and the coordinates are expressed in the units of $c/\omega$, $c$ is the speed of light.

For the structures with slowly varying periods, the eigenmodes of the problem~(\ref{Helmholtz}) can be introduced
by applying the Bloch theorem for periodic systems and the corresponding Bloch-wave formalism~\cite{Yeh:1979-742:JOSA}.
In this approach, the electric field corresponding to the eigenmode with the propagation constant $k_y$ in the unit cell
with the width $\Lambda$ can be represented as,
\bea \label{eq_eigenmode}
 E_\Lambda(z,y)=A_{\Lambda}U_{\Lambda}(z)\exp \left[-i(K_b^\Lambda z + k_y y)\right],
\eea
where $k_y$ and $K_b$ are normalized to $\omega/c$, the amplitude $A_\Lambda$ varies slow across the structure,
$U_\Lambda$ is the Bloch function, and $K_b^\Lambda$ is the Bloch number of the periodic structure
with the unit cell size $\Lambda$, which we analyze systematically in Sec.~\ref{sec_periodic}.

We study propagation of the Gaussian beams with the electric field at $y=0$ defined as
\bea \label{eq_Gauss}
   E_x(z,y=0)=\exp[-(z-z_0)^2/h]E(z)=\psi(z)E(z),
\eea
where $E(z)$ is the eigenmode of the Eq.~(\ref{eq_eigenmode}) with the propagation constant $k_y=k_y^0$, which will determine propagation direction; $h$ is the beam width, which we consider to be significantly large, i.e. $h \gg \; <\Lambda>$. The field in the structure can be represented as a superposition of eigenmodes. As
long as the Gaussian beam is wide with respect to the unit cell, the spectrum of the contributing eigenmodes is narrow, and it is centered near the point $k_y^0$~\cite{Yeh:1979-742:JOSA}. Such beam can be treated within the paraxial approximation. The beam propagation is mainly described by the modes with $k_y \simeq k_y^0$ and the tangent of the beam angle is defined by $(\partial k_y/\partial K_b)_{\mid k_y=k_y^0, \Lambda=\Lambda_0}$, where $\Lambda_0$ corresponds to the beam launching point. An additional beam tilt can be introduced in more general case of complex $\psi(z)$.

\pict[1]{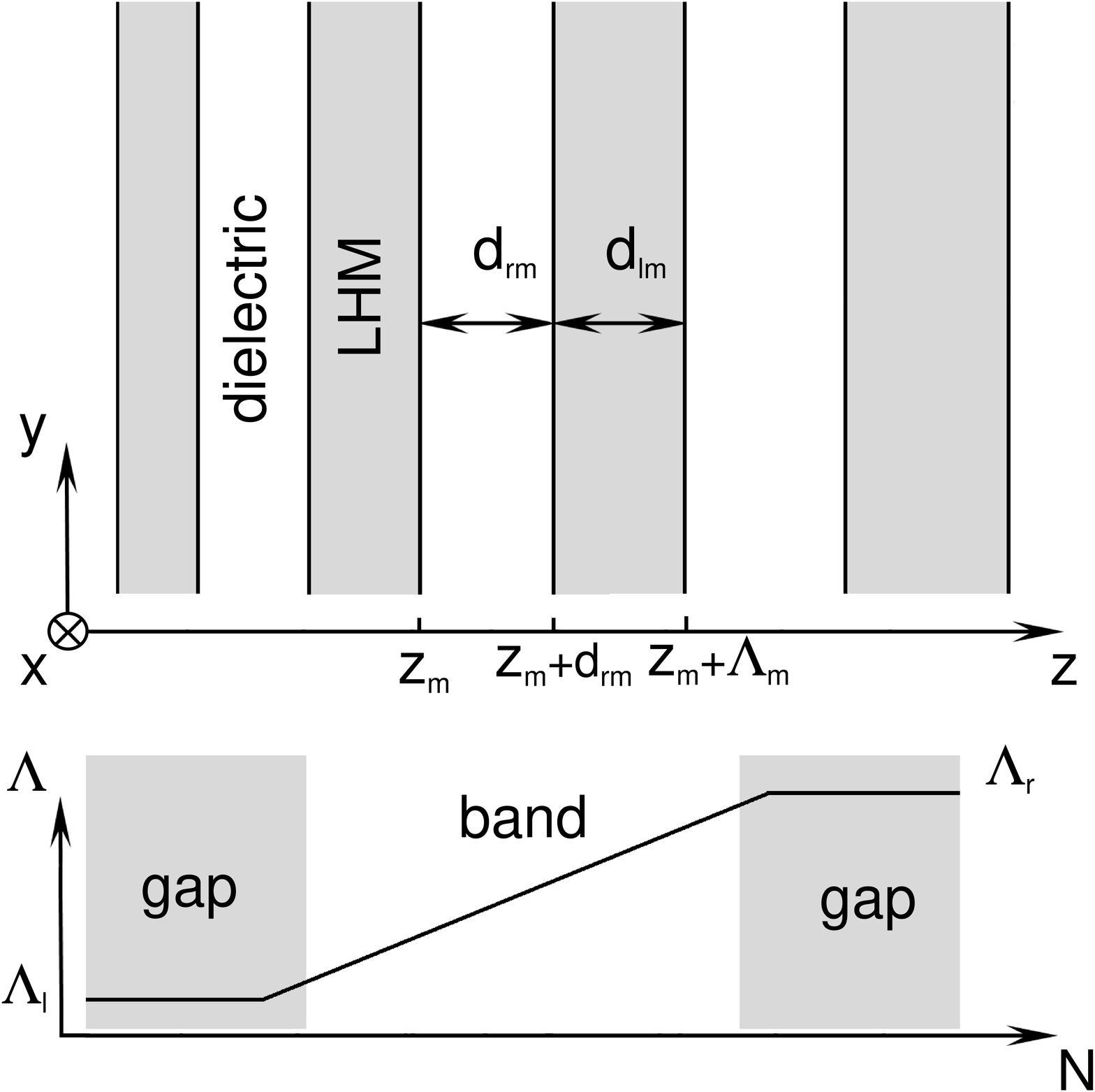}{fig_stacks}{Top: Schematic of the chirped layered structure with a linearly growing period $\Lambda_m$. Shaded slabs correspond to metamaterial layers with the width $d_{lm}$
separated by dielectric layers with the width $d_{rm}$. Bottom: Schematic of the period variation across the structure. Structure with a linearly changing period is embedded into semi-infinite
periodic structures with the periods $\Lambda_{\it l}$ and
$\Lambda_{\it r}$, respectively.}

\section{Periodic Structure: Bandgap Properties}
\label{sec_periodic}

As the first step of our analysis, we study the periodic structure with the period $\Lambda$ without
a chirp. In this case Eq.~(\ref{Helmholtz}) has periodic coefficients, and we can apply the Bloch
theorem~\cite{Yeh:1988:Optical_Waves} and present the electric field in the form,
\bea \label{eq_electric_field}
  E_x(z,y)=U(z)\exp[-i(K_b z + k_y y)],
\eea
where $K_b$ is the Bloch number, and the field envelope $U(z)$ is a periodic function with
the period $\Lambda$, so that $U(z+\Lambda)=U(z)$.  In the $m$-th pair of layers, the
Bloch waves are presented as,
\bea \label{eq_Bloch_envelope}
U_{l,r}(z)=\left[a_{l,r}exp\left(-i k_{zl,zr}(z-m\Lambda)\right)
\right.+ \\ \left.
b_{l,r}exp\left(i k_{zl,zr}(z-m\Lambda)\right)\right]
\exp(iK_b(z-m\Lambda)),
\nonumber \eea
where the indices {\it l} and {\it r} correspond to the left- and
right-handed slabs, respectively, and the amplitudes $a_{l,r}$ and
$b_{l,r}$ are found from the boundary conditions at the interfaces
separating the metamaterial and dielectric layers (see,
e.g., Refs.~\cite{Yeh:1988:Optical_Waves,Born:1964:Principles_of_optics,Yeh:1979-742:JOSA}).
The Bloch wavenumber $K_b$ is defined from the dispersion relation,
\bea \label{eq_Bloch_wavevector_dispersion}
2\cos(K_b\Lambda) =
    2\cos(k_{zr} d_r) \cos(k_{zl}d_l) - \\
    \left(
        \frac{k_{zl}\mu_r}{k_{zr}\mu_l}
        +\frac{k_{zr}\mu_l}{k_{zl}\mu_r}
    \right)
    \sin(k_{zr}d_r)\sin(k_{zl}d_l),
\eea
where $k_{zl,zr}=\mp \sqrt{n_{l,r}^2-k_y^2}$ and $k_y$ is the
propagation constant along the $y$ axis.

\begin{figure}
\includegraphics[width=1\columnwidth]{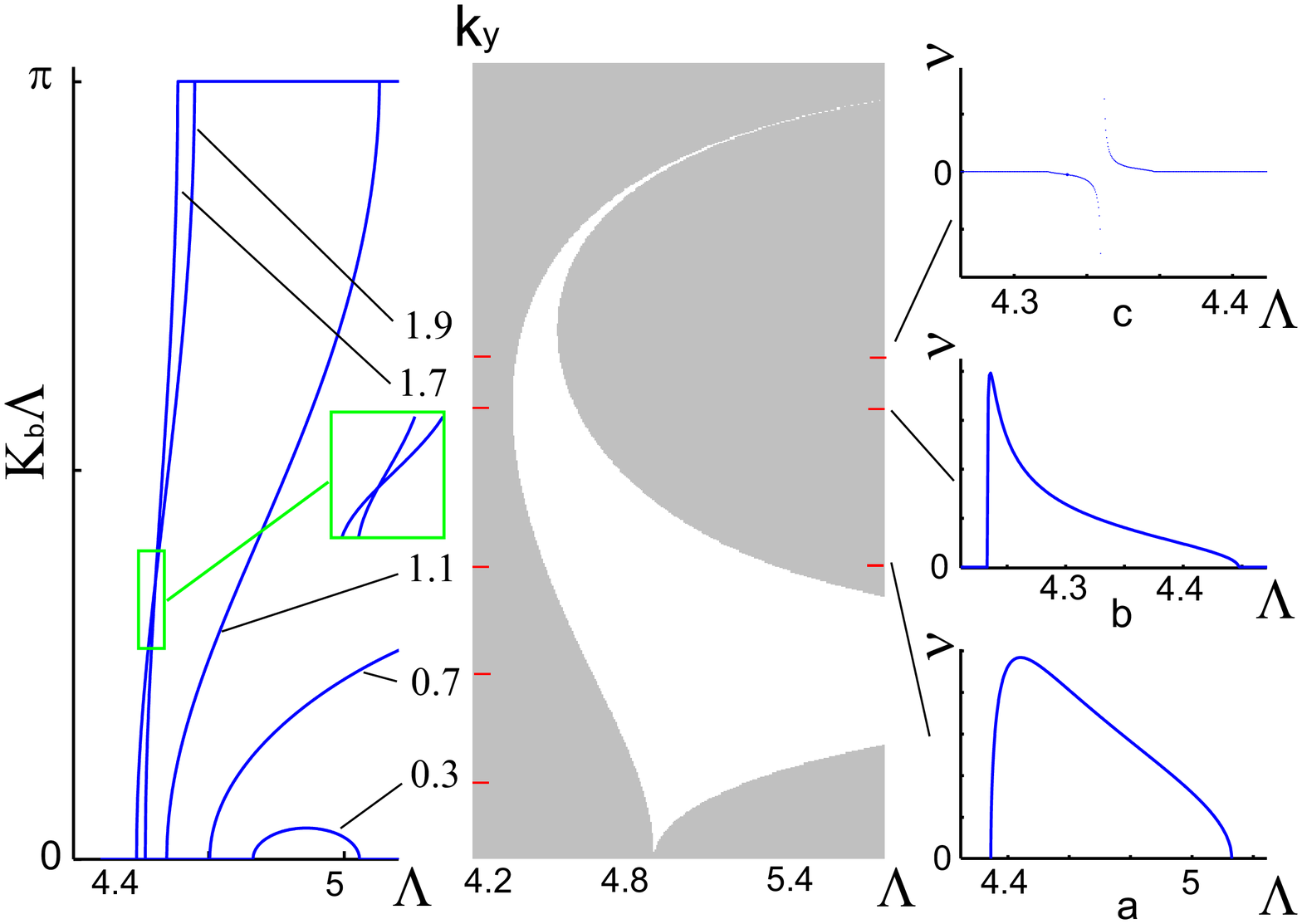}
\caption{(Color online) Left: Bloch-wave phase incursion in the transmission regime for different
values of the propagation constant $k_y$. Inset shows a magnified part of the plot
with the mode crossing. Middle: bandgap diagram for $\varepsilon_r
=\mu_r=1$, $a/b=2$ and $\varepsilon_l=-5$, $\mu_l=-0.8$; low-order transmission
resonance is shown. Right: group velocity $\partial k_y / \partial K_b$
for different values of the propagation constant.}
\label{fig_bandgap1}
\end{figure}

According to Eq.~(\ref{eq_electric_field}), the waves can propagate in the structure when $K_b$ is real.
In Fig.~\ref{fig_bandgap1}(middle), we plot the bandgap diagram of the layered structure on the plane $(\Lambda,k_y)$.
Here we assume that the dielectric layer is air, $\varepsilon_r=\mu_r=1$, and the dielectric layers are two times
thicker than the metamaterial layers, i.e.  $d_{rm}/d_{lm}=2$, and this ratio is preserved
across the structure. We choose the metamaterial with the parameters $\varepsilon_l=-5$ and $\mu_l=-0.8$.

The bandgap properties of one-dimensional periodic structures with metamaterials
have been studied in several papers~\cite{Li:2003-083901:PRL,Wu:2003-235103:PRB}, including the
case of the zero-index averaged refractive index~\cite{Shadrivov:2005-193903:PRL}. It was shown that for the
structures with $\int \limits_\Lambda n(z) dz= 0$ the bandgap spectrum includes transmission resonances which shrink
into infinitesimally thin lines into a complete bandgap when the widths of the slabs coincide
~\cite{Shadrivov:2005-193903:PRL}. For the normal incidence
(i.e. when $k_y=0$), the transmission is observed only when $n_r d_r=n_l d_l=
\pi q$, where $q$ is integer.

We study the Bloch wave phase incursion $K_b\Lambda$ across the
structure for different values of the propagation constant $k_y$,
see Fig.~\ref{fig_bandgap1}(left). For small values
of the propagation constant (e.g., $k_y<0.5$),  the low-order
transmission resonances have zero phase incursion, in contrast to
the case of conventional dielectric Bragg gratings. The maximum of
the phase is accumulated in the middle of the transmission band, and
its amplitude increases with the increase of the band order, reaching the value
of $\pi$ in the n-th order transmission resonance. A growth of the propagation constant leads to the band coupling
($0.5\lesssim k_y\lesssim 1$), not shown here. The phase incursion
in a new coupled broadband region is approaching $\pi$. Further
increase of the propagation constant leads to the narrowing of the
transmission regions and to a rapid increase of the phase
across the band to the value of $\pi$. On the bandgap diagram, we
observe two {\it turning points} at $k_y \simeq 1.7$ and $k_y \simeq 2$, (marked points 1 and 2 in the inset of Fig.~\ref{fig_bandgap2}(middle)). At these points the modes with close $k_y$ have the same value of the Bloch wavenumber near the band edge. We trace the behavior of the band edges $\zeta_-(k_y)$ and $\zeta_+(k_y)$ with a change of the propagation constant $k_y$. For $k_y \simeq 1.7$, the left band edge $\zeta_{-}$ changes the type of its monotonicity. Consequently, in this region the left ($\zeta_-$) and right ($\zeta_+$) band edges have different slopes, i.e.
$(\partial \zeta_-/\partial k_y) (\partial \zeta_+/\partial k_y) <0$, as shown in the inset of Fig.~\ref{fig_bandgap2}. The latter, with consideration that phase is growing from 0 to $\pi$ in the transmission region, leads to the {\em crossing} of the corresponding modes (see inset in Fig.~\ref{fig_bandgap1}). Finally, for  $k_y \simeq 2$, the right band edge $\zeta_+$ also changes the sign of its slope, and $(\partial \zeta_-/\partial k_y) (\partial \zeta_+/\partial k_y) $ becomes positive. We note that for small values of $k_y$, the first-order band edges also have different signs of their slopes but a new type of behavior cannot be observed because the phase incursion vanishes in the band for $k_y<0.5$. Also, in the conventional Bragg gratings both band boundaries have the same slope and thus the mode crossing does not occur.

Crossing of two modes means that the effective spatial velocity defined as $\nu=\partial k_y / \partial K_b$, has different signs at different sides of the crossing point, and it becomes singular at the point itself. As long as the velocity is related to the energy flow, we observe different directions of the energy flow in the $y$-direction at different edges of the band, as will be discussed in more details in Sec.~\ref{sec_geometry} and Sec.~\ref{sec_field}. We plot the velocity for different values of the propagation constant $k_y$, see Figs.~\ref{fig_bandgap1}(a-c). For $k_y$ outside the region, Fig.~\ref{fig_bandgap1}(a), the velocity does not change its sign in the band, thus meaning that energy flow has the same direction at any point of the band. Similar type of the beam evolution can be found in the stacks of conventional dielectric layers. Further increase of $k_y$ leads to more asymmetric velocity profile and, finally, at $k_y \simeq 1.7$ [see Fig.~\ref{fig_bandgap1}(b)], the velocity becomes infinite at the left band edge. In Sec.~\ref{sec_geometry} below we demonstrate that this case corresponds to the energy flow vanishing in the $y$-direction. When the propagation constant $k_y$ is inside the region [see Fig.~\ref{fig_bandgap1}(c)], the velocity profile has the second-order discontinuity with the sign changed inside the band. Consequently, the energy flow in the $y$-direction also changes its sign within the transmission band. The singularity point in the spatial velocity profile moves from the left band edge to the right edge with the growth of the propagation constant from $k_y \simeq 1.7$ to $k_y \simeq 2$, (see {\em turning points} 1 and 2 in Fig.~\ref{fig_bandgap2}). Finally,  for the propagation constants $k_y>2$, the velocity profile becomes continuous and negative, indicating that the waves are backward in the
whole band.

Below, we consider the simplest case of a linear chirp of the structure, $\Lambda_{i+1}=\Lambda_i+\delta \Lambda$ with $\delta \Lambda \ll \Lambda$, see Fig.~\rpict{fig_stacks}. For $\delta\Lambda \rightarrow 0$, i.e. an adiabatic change of the unit cell size across the structure, we implement the geometric optics approximation.

\section{Geometric Optics Approximation}
\label{sec_geometry}

We study the evolution of the Poynting vector by employing an analogy with a homogeneous medium
having gradually changing refraction index
~\cite{Born:1964:Principles_of_optics,Landau:2001:Theoretical_Physics}. We consider the time-averaged Poynting vector,
\bea \label{eq_time_average_poyting_vector} {\bf S}=\frac{c}{8\pi}
\textrm{Re}[{\bf E\times H^*}]= \; \; \; \; \; \; \;\\
=-\frac{c}{8\pi \mu}\textrm{Re} \; i\left[{\bf
e_z}E_x\frac{\partial {E_x}^*}{\partial z}+{\bf
e_y}E_x\frac{\partial {E_x}^*}{\partial y} \right], \nonumber \eea
where $E_x$ is given by Eqs.~(\ref{eq_Gauss}), ${\bf e_y}$ and ${\bf
e_z}$ are the unit vectors. The
time-averaged electromagnetic energy density is defined as
\bea \label{eq_time_average_energy_dencity}
w = \frac{1}{16 \pi} \textrm{Re}(\varepsilon {\bf E E^*}+\mu {\bf H H^*})= \; \; \; \; \;\\
=\frac{1}{16 \pi \mu}\textrm{Re}\left(\varepsilon \mu
|E_x|^2+\left|\frac{\partial E_x}{\partial
z}\right|^2+\left|\frac{\partial E_x}{\partial y}\right|^2\right),
\nonumber \eea

In a homogeneous medium, the time-averaged Poynting vector is tangential to the beam trajectory. It changes gradually with continuously varying refractive index, thus defining a geometric ray. In our case, the Poynting vector changes its direction stepwise within each unit cell, since the refractive indices of the slabs comprising the unit cell have the opposite signs. To describe an average behavior of the energy flow on larger scale, we consider the Poynting vector and energy density averaged over the unit cell, $<{\bf
S}>=\frac{1}{\Lambda}\int \limits_0^{\Lambda}{\bf S}dz$ and
energy density $<w>=\frac{1}{\Lambda}\int \limits_0^{\Lambda}w dz$, respectively.

We introduce the average velocity of the energy flow~\cite{Yeh:1979-742:JOSA,Yeh:1988:Optical_Waves}:
\bea \label{eq_energy_velocity} {\bf v_e}=\frac{<{\bf
S}>}{<w>}, \eea

As long as the width of the unit cell changes adiabatically across
the structure, the average velocity of the energy flow ${\bf v_e}$
changes quasi-continuously. Therefore, we can define the rays in the space $(\Lambda,y)$ (we note here that due to the linear chirp of the structure, the spatial coordinate $z_m$ is a linear function of $\Lambda$) as trajectories along which the average energy $<{\bf S}>$ is directed. Thus, the ray equation is defined as follows:
\bea \label{eq_ray}
  \frac{d{\bf r}}{dt} = {\bf v_e},
\eea
where ${\bf r}={\bf e_y}y+{\bf e_z} \Lambda$ is the radius-vector
describing the ray.

In Ref.~\cite{Yeh:1979-742:JOSA} it was shown that for a paraxial beam the average velocity of
the energy flow coincides with the group velocity defined as
\bea \label{eq_group_velocity}
  {\bf v_{e}}={\bf v_g}=\nabla_{\bf K} \omega = \frac{\partial \omega}{\partial {\bf K}},
\eea
where ${\bf K}={\bf e_y}k_y+{\bf e_z}K_b$ is the wave vector, and $\omega$ is the angular frequency.

For a fixed frequency $\omega$ of monochromatic beam we obtain the dispersion relation  $k_y=f(K_b)$, (see Eq.~\ref{eq_Bloch_wavevector_dispersion}), corresponding to the curve in the space $(k_y,K_b)$ described by the wave vector ${\bf K}$, which is analogous to the equifrequency surface for anisotropic media~\cite{Kravcov:1980:Geometric_optics_nhm_media}. Consequently, using Eqs.~(\ref{eq_energy_velocity}), (\ref{eq_ray}), and (\ref{eq_group_velocity}) we obtain ${\bf v_g}={\bf v_e}$ and $<{\bf S}>$ are normal to the curve $k_y=f(K_b)$. The latter condition is described by the equation,
\bea \label{eq_normal}
  \frac{<S_y>}{<S_z>}=-\left(\frac{\partial k_y}{\partial K_b} \right)^{-1}\equiv-\frac{1}{\nu},
\eea
where $<S_y>$ and $<S_z>$ are the $y$ and $z$ components of the Poynting vector $<{\bf S}>$, $\nu$
is the spatial velocity of the ray.

Relation~(\ref{eq_normal}) between the Poynting vector components and velocity
$\nu$ shows that when $\nu \rightarrow \infty$, $<S_y>=0$, and
the total energy flows along the $z$-direction. On the other hand, when $\nu = 0$, the energy flows along the structure ($<S_z>=0$).

Using Eqs.~(\ref{eq_energy_velocity}) and (\ref{eq_normal}), we rewrite Eq.~(\ref{eq_ray}) as follows:
\bea \label{eq_ray_projections}
  \frac{dy}{dt}=\frac{<S_y>}{<w>} \; \; \; \; \; \; \; \; \; \; \; \; \; \; \; \; \\
  \frac{d \Lambda}{dt}=\frac{<S_z>}{<w>}=\frac{-<S_y>\nu}{<w>}. \nonumber
\eea
Now we can derive the ray trajectory equation in the space $(\Lambda,y)$
\bea \label{eq_motion}
  \frac{d \Lambda}{dy}=\mp \frac{\partial k_y}{\partial K_b},
\eea
where $\mp$ correspond to the forward and backward propagating waves, respectively.

We note that the motion equation (\ref{eq_motion}) can also be  derived from the Fermat principle. We multiply
Eq.~(\ref{eq_ray}) by ${\bf K}$ and using Eq.~(\ref{eq_group_velocity}), we write the Fermat principle
in the form~\cite{Kravcov:1980:Geometric_optics_nhm_media}:
 \be \label{eq_Fermat_principle}
   \delta \int{\bf K}d{\bf r}=0б,
 \ee
where $d{\bf r}={\bf e_y}dy+{\bf e_z}d\Lambda$ is the unit vector along the beam trajectory.
The Fermat principle is the principle of least action with the coordinate $y$ playing the role of time. The Lagrange function $L=\pm K_b(d\Lambda/dy)+k_y$ is defined accurate to the full $y$-derivative of ${\Lambda}$. As a result, the unambiguity in the Bloch wavenumber, $K_b=K_b^{(0)}+2\pi m / \Lambda$ does not affect the ray behavior. The Hamilton equations for the rays can be derived in the form~\cite{Landau:2001:Theoretical_Physics}:
\bea \label{eq_Hamilton} \frac{d\Lambda}{dy}=\mp\frac{\partial
k_y}{\partial K_b}=\mp \nu \; \; \; \; \; \;
\frac{dK_b}{dy}=\pm\frac{\partial k_y}{\partial \Lambda}, \eea

We remind now that $k_y$ remains constant across the structure,
consequently $\partial k_y/\partial \Lambda=0$, meaning that the
problem is invariant to translation along the layers. Hence, we
consider only the first equation, which can be also derived from
general suggestions, assuming that $k_y$ refers to the "energy",
being preserved in the structure and $\partial k_y/\partial K_b$
being the "group velocity" of the paraxial beam.

\begin{figure}
\includegraphics[width=1\columnwidth]{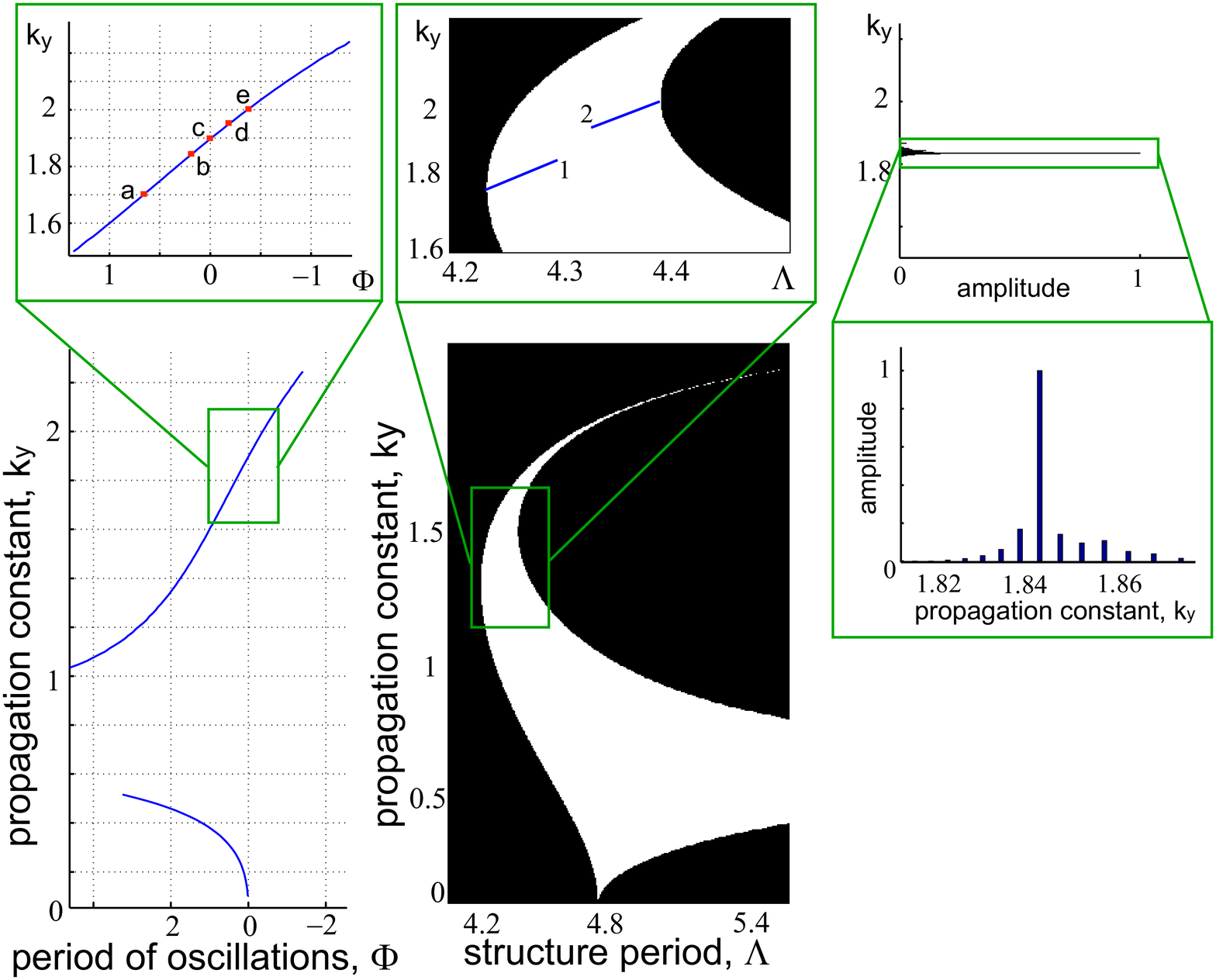}
\caption{(Color online) Left: period of beam oscillations. Inset shows the region of the sign change. Middle: bandgap
diagram. Inset shows the turning points. Right: spectrum of the beam centered at the point $k_y^0=1.843$}\label{fig_bandgap2}
\end{figure}

The first equation of the system~(\ref{eq_Hamilton}) describes diffractionless motion of a beam with a narrow spectrum centered near the point $k_y^0$. The beam motion is equivalent to the motion of an effective
particle in a one-dimensional potential $W=[\nu^2(\Lambda)/2]$ between left (${\it \zeta_-}$) and right (${\it \zeta_+}$) band edges. As long as the spectrum of allowed energies near the point $k_y^0$ is quasi-continuous for infinite structures with adiabatically changing width $\Lambda$, the motion remains periodic with the period of
oscillations $\Phi$ found as,
\bea
  \Phi(k_y) = 2\int\limits_{\it \zeta_-}^{\it \zeta_+}
  \frac{d\Lambda}{\nu(\Lambda,k_y)},
\eea
where $\zeta_-$ and $\zeta_+$ refer to the band edges, $\nu$ is defined by Eq.~(\ref{eq_normal}). In the gap regions, the field decays exponentially across the structure so that the motion is prohibited, and the spatial velocity vanishes at the band edges. Consequently, the expression under the integral has singular points at ${\it \zeta_\pm}$,
and the integration fails. To resolve this problem, we introduce a new variable $\xi$,
\bea
  \Lambda = \frac{{\it \zeta_-}+{\it \zeta_+}}{2}+\frac{{\it \zeta_-}-{\it \zeta_+}}{2}\cos(\xi),
\eea
and obtain
\bea \label{eq_oscillation_period}
  \Phi = 2\int\limits_0^\pi \frac{({\it \zeta_+}-{\it \zeta_-})}{2}
  \frac{d\xi}{\nu(\xi,k_y)}.
\eea
The expression under the integral is finite within the integration boundaries, hence the new integral is converging.
Using the integral~(\ref{eq_oscillation_period}) we calculate the dependence of the period on the propagation constant $k_y$, see Fig.~\ref{fig_bandgap2}(left).

We observe that with a growth of $k_y$ the oscillation period grows as well, this is explained by broadening of the band described in Sec.~\ref{sec_structure}. A further increase of $k_y$ leads to the beam narrowing and decrease of the oscillation period $\Phi$. For $1.7 < k_y < 2$, i.e between the turning points, the period vanishes meaning that the corresponding mode does not transfer energy along the structure. A further increase of $k_y$ leads to negative values for the period of oscillations, which corresponds to the total energy flow in the direction opposite to the propagation constant ${\bf k_y}$.

\begin{figure}
\includegraphics[width=\columnwidth]{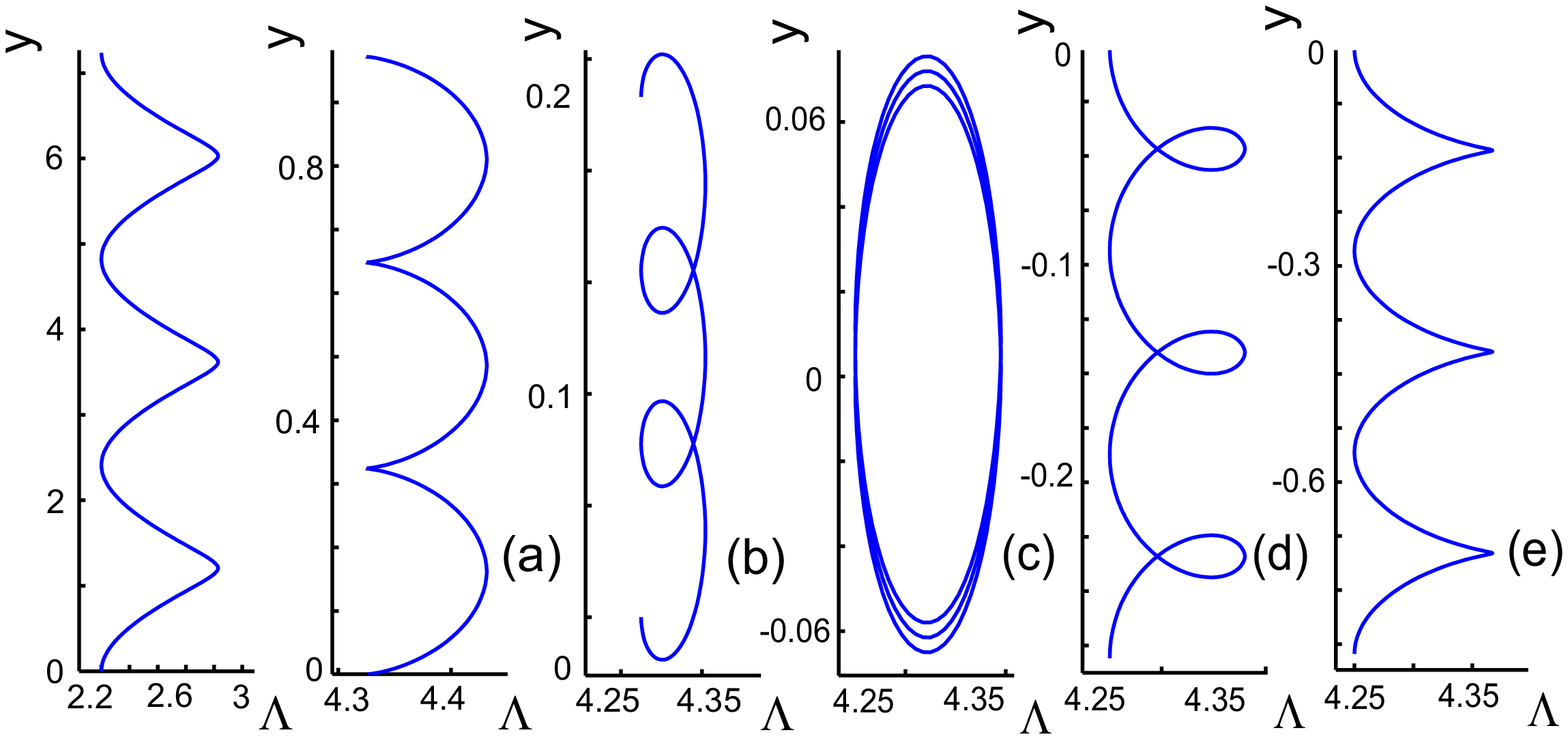}
\caption{(Color online) Ray trajectories calculated in the geometric optics approximation. Left: conventional
dielectric structures, for $\mu_a=\varepsilon_a=1$ and $\mu_b=1$; $\varepsilon_b=10$. (a-e) Structures with metamaterials for $k_y^0=1.7, \; 1.85, \; 1.895, \; 1.95, \; 2,$ respectively.}
\label{trajectory}
\end{figure}

Using Eq.~(\ref{eq_Hamilton}) and Eq.~(\ref{eq_oscillation_period}), we calculate the beam trajectories. First of all, we verify our results for the well-known structures with conventional dielectric slabs, and the trajectory corresponding to this case is shown in Fig.~\ref{trajectory}(left). We assume that a dielectric layer with the width $b$ has the parameters $\varepsilon_b=10$ and $\mu_b=1$, and the layers are separated by vacuum of the width $a$, so that $a/b=2$. We reveal a close correspondence between our results and the trajectories calculated by means of the geometric optics in Ref.~\cite{Wilkinson:2002-056616:PRE}. The beam is reflected from both band boundaries,
where the spatial velocity vanishes. The total energy flow along the $y$-direction is positive, and it preserves its sign during oscillations. For the structures with metamaterials, the beam with the spectrum centered near the propagation constants $k_y$ outside of the region between turning points (points 1 and 2 in Fig.~\ref{fig_bandgap2}) experiences the same reflection from the boundaries of the Brillouin zone with the total energy flow along the structure in the positive direction, for $k_y<1.7$, and in opposite (negative) direction, for $k_y>2$.

Next, we study the beam propagation for the values of $k_y$ inside the region between the turning points. The trajectory calculated near the first turning point (see Fig.~\ref{trajectory}) has a peculiarity near the left band
edge. Corresponding period of oscillations is shown in Fig.~\ref{fig_bandgap2}(left). Near the left band edge the group velocity becomes infinite, thus energy flow along the layers vanishes, and the energy flow across the structure changes its sign. Hence a vortex structure is formed at this point. The observed spike is formed by narrowing of the trajectory near the left band edge with increase of propagation constant until it reaches the first turning point. Further increase of propagation constant, as was described in Sec.~\ref{sec_structure}, corresponds to a shift of the singular point of the group velocity further into the band region. The trajectories corresponding to this case are shown in Figs.~\ref{trajectory}(a-e). Since the group velocity changes its sign before and after the singular point, the trajectory crosses itself and a beam curl is formed. For $k_y^0=1.85$ we observe a curl near the left band edge
of the structure, see Fig.~\ref{trajectory}(b). Further increase of the propagation constant leads to the increase of the curl's size, and decrease of the period, see Fig.~\ref{fig_bandgap2}. Finally, at $k_y^0 \simeq 1.895$ the curl reaches the other transmission band edge forming a practically closed trajectory with almost zero period of oscillations, see Fig.~\ref{trajectory}(c). Note that near the left band edge the energy flow in the $y$-direction is
negative but remains positive near the right band edge. Further increase of the wavenumber results in the opposite process with the curl forming near the right band edge, see Fig.~\ref{trajectory}(d). For the propagation constants $k_y$ corresponding to the second turning point the curl vanishes, with a spike forming near the right band edge, and the total energy flow along the structure is in the opposite direction to the propagation constant. For $k_y^0>2$, the spike vanishes, and smooth reflection from the right band edge is formed. The trajectory again becomes continuous and without any ambiguities.

\section{Numerical Simulations}
\label{sec_field}

For the finite structures with non-adiabatic change of the unit-cell size, the approximation of geometric optics is not valid, however we can still apply the Bloch-wave formalism for describing the beam propagation, including the beam diffraction and formation of interference patterns. To find the field distribution in the structure, we calculate the eigenmodes $E_i(z)$ and eigenvalues $k_y^i$ of a finite stack of layers. There are several known  approaches to solve this problem. The first approach is based on the representation of the field in each slab in the form of the counter-propagating waves. Boundary conditions between the layers define the ratio of the corresponding wave amplitudes in the slabs. These amplitudes can be represented in terms of either transfer matrix or scattering matrix formalisms~\cite{Yeh:1988:Optical_Waves, Ko:1988-1863:JOSA}. The transfer matrix approach can be used for calculating the eigenmodes and eigenvalues for the structures with a small number of periods, and it is not suitable for calculating evanescent modes in long enough structures. The scattering matrix approach provides a better convergence, but it also cannot be applied for calculating evanescent modes in the structures with more than approximately 100 slabs. Another approach is based on the discretization of the wave equation across the structure, and we have used it previously to find the eigenvalues of the discrete problem~\cite{our_oe}. However, this approach is computationally intensive, and it is not practical for calculating the transmission of very long structures.

Here we implement another approach based on the Bloch wave formalism. In Sec.~\ref{sec_structure} we have already demonstrated that for the slowly varying period of the structure (i.e., $\delta \Lambda \ll \Lambda$) the eigenmodes can be described by Eq.~(\ref{eq_eigenmode}) with the amplitudes $A_{\Lambda}$ and wavenumbers $k_y$. We assume that the structure with a linearly changing period is placed between two semi-infinite periodic structures, with the period coinciding with that of the adjacent layers of the structure, see Fig.~\rpict{fig_stacks}.
In each unit cell the field can be presented as a superposition of the forward and backward propagating Bloch waves,
\bea \label{eq_eigenfield}
E_x^{\Lambda}(y,z)=A_{\Lambda}U_+(z)\exp(-i(K_bz+k_yy))+\\
B_{\Lambda}U_-(z)\exp(-i(-K_bz+k_yy)), \; \; \; \; \; \; \; \; \; \;
\nonumber \eea
where $A_{\Lambda}$ and $B_{\Lambda}$ are the slowly varying amplitudes, $U_{\pm}$ and $K_b$ are defined by
Eqs. ~(\ref{eq_Bloch_envelope}) and ~(\ref{eq_Bloch_wavevector_dispersion}), respectively.

\pict[1]{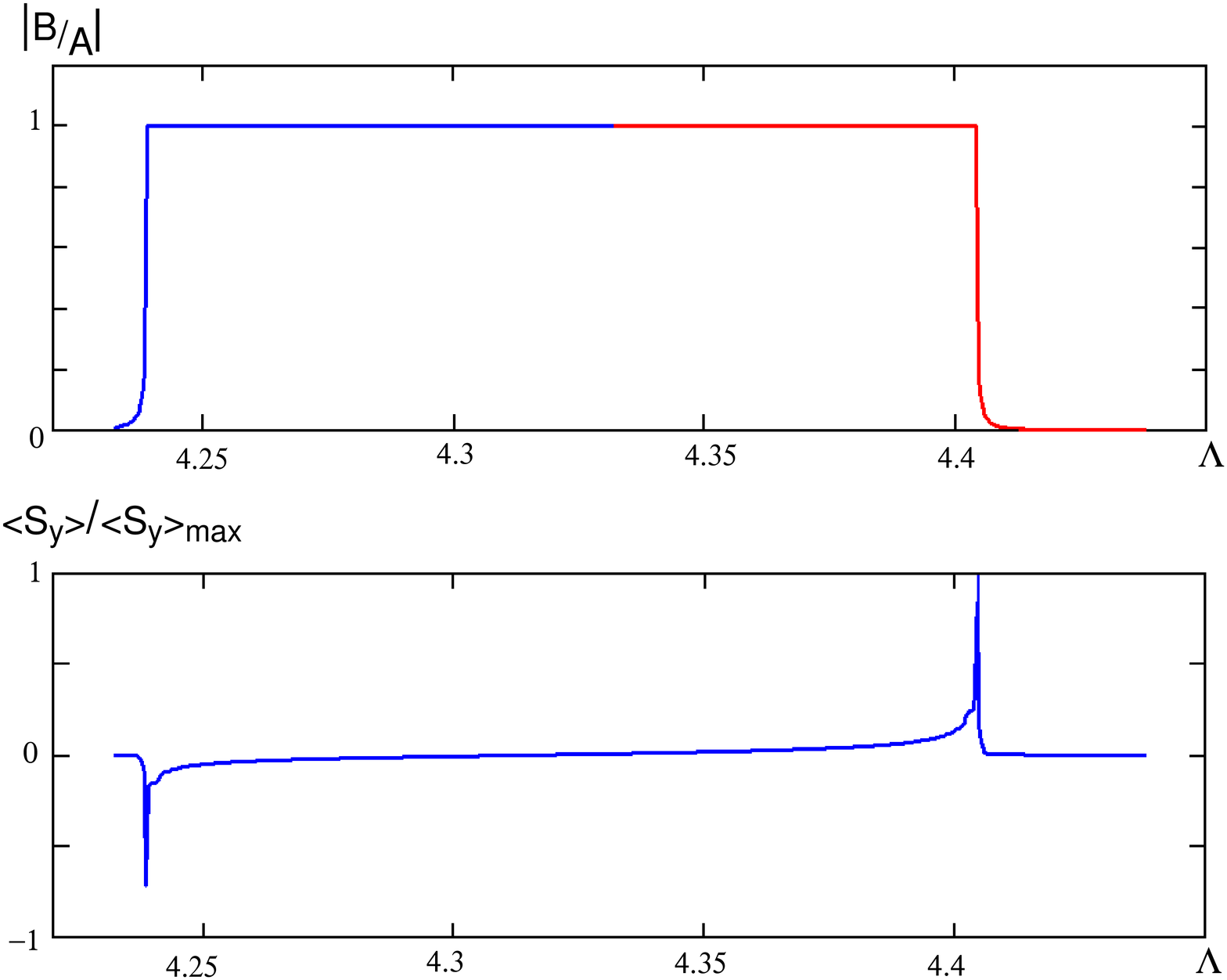}{fig_amplitudes}{(Color online) Top: ratio between the amplitudes of
the counter-propagating Bloch waves across the structure calculated for $k_y = 1.838$.
Bottom:  energy flow averaged by the unit cell across
the structure, for $k_y = 1.838$.}

In the left semi-infinite structure, the fields decay exponentially, and we choose the amplitudes in the first layer as $A_{\Lambda_l} = 1$ and $B_{\Lambda_l}=0$. In the right semi-infinite structure the forward wave vanishes, and this means that $A_{\Lambda_r}=0$ and $B_{\Lambda_r}=1$. We calculate the amplitudes
in the adjacent unit cells using the boundary conditions. Repeating this procedure we find the amplitudes in the whole structure (for more details, see Appendix A). We note that $K_b$ is real in the band, hence the Bloch waves are propagating, and the modes do not grow exponentially. In a small number of layers adjacent to the band, $K_b$ is imaginary and, consequently, the amplitudes grow exponentially. We choose relatively small number of unit cells with the parameters corresponding to the bandgap, so that the amplitudes $A_\Lambda$ and $B_\Lambda$ remain reasonable across the structure. We note that the absolute value of the ratio of the amplitudes of the counter-propagating waves in the band is unity, so that the energy is localized in the transverse direction inside the structure. We trace the dynamics of the Bloch wave amplitudes from left and right boundaries separately and compare the amplitudes in the middle of the band region, see Fig.~\rpict{fig_amplitudes}. For the eigenmodes with the eigenvalues $k_y$, the amplitudes of the waves at the left and right edges of the structures become linearly dependent, i.e.,
\bea
 \det\left.\left(
  \begin{array}{cc}
    1 & 1 \\
    B_l/A_l & B_r/A_r
  \end{array}
  \right) \right|_{\Lambda=(\Lambda_r+\Lambda_l)/2} = 0.
\eea
Using the procedure described above, we find the propagation constants $k_y$ corresponding to the eigenvalues and eigenmodes of the problem. Then, the initial field distribution is represented as a superposition of the eigenmodes, and the full structure of the fields can be retrieved.

We calculate the eigenmodes for the structure consisting of 2100 unit-cells with the width changing linearly from the value $\Lambda_l = 4.23$ to the value $\Lambda_r=4.44$, and $\delta \Lambda \simeq 2\times10^{-4}$. We note that $<\Lambda>=(\Lambda_r+\Lambda_l)/2\simeq2/3\lambda$, where $\lambda$ is a free space wavelength. The width of the whole structure is approximately $1500\lambda$, which is, e.g., about few millimeters in infrared regime. We have chosen the initial field distribution in the form of Gaussian beam, see Eq.~(\ref{eq_Gauss}), launched near the right band boundary of the structure, see Fig.~\rpict{fig_field_distribution}. The width of the beam is $a = 20 (\Lambda_l+\Lambda_r)/2$, i.e with about 20 unit-cells exited, $k_y^0=1.838$. We decompose the initial field distribution into the superposition of eigenmodes of the structure by the least-squares method. The spectrum of the excited eigenmodes is shown in Fig.~\ref{fig_bandgap2}(right). The spectrum is narrow, and it is centered near $k_y=1.838$. The spectrum is to some accuracy equidistant, meaning that the field restores its shape after the distance $\Phi=2\pi/\delta k_y$. We plot the field amplitude averaged over the unit cell. The beam curling predicted in the framework of the geometric optics is clearly observed. At the beam self-crossing point we observe an interference pattern created by the forward and backward propagating waves. The field distribution is restored almost completely after the first period of oscillations.

\pict[1]{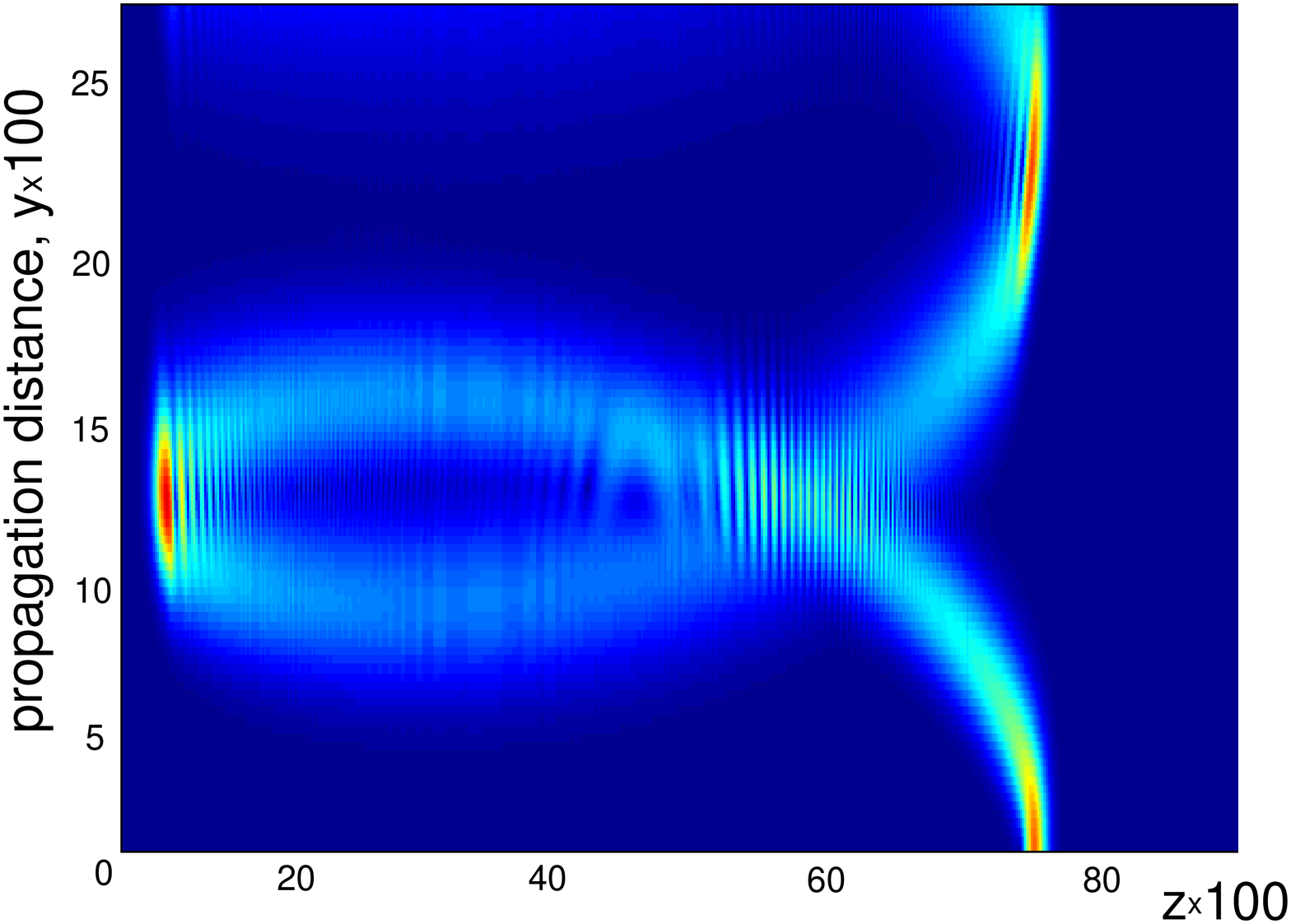}{fig_field_distribution}{(Color online) Distribution of the field amplitude in the structure, for the beam with the width $a = 20 \Lambda$, at $k_y^0=1.838$. Oscillation period is about $23,5\times100$.}

Now we calculate the distribution and direction of the Poynting vector. In Fig.~\rpict{fig_Poyting_distribution} we plot the Poynting vector averaged across the unit cells. It is clearly seen that near the left boundary the direction of the Poynting vector along $y$-axis is opposite to the direction near the right boundary. At the self-crossing point corresponding to singularity of the group velocity, we observe that the energy flow vanishes in the $y$-direction.

\pict[1]{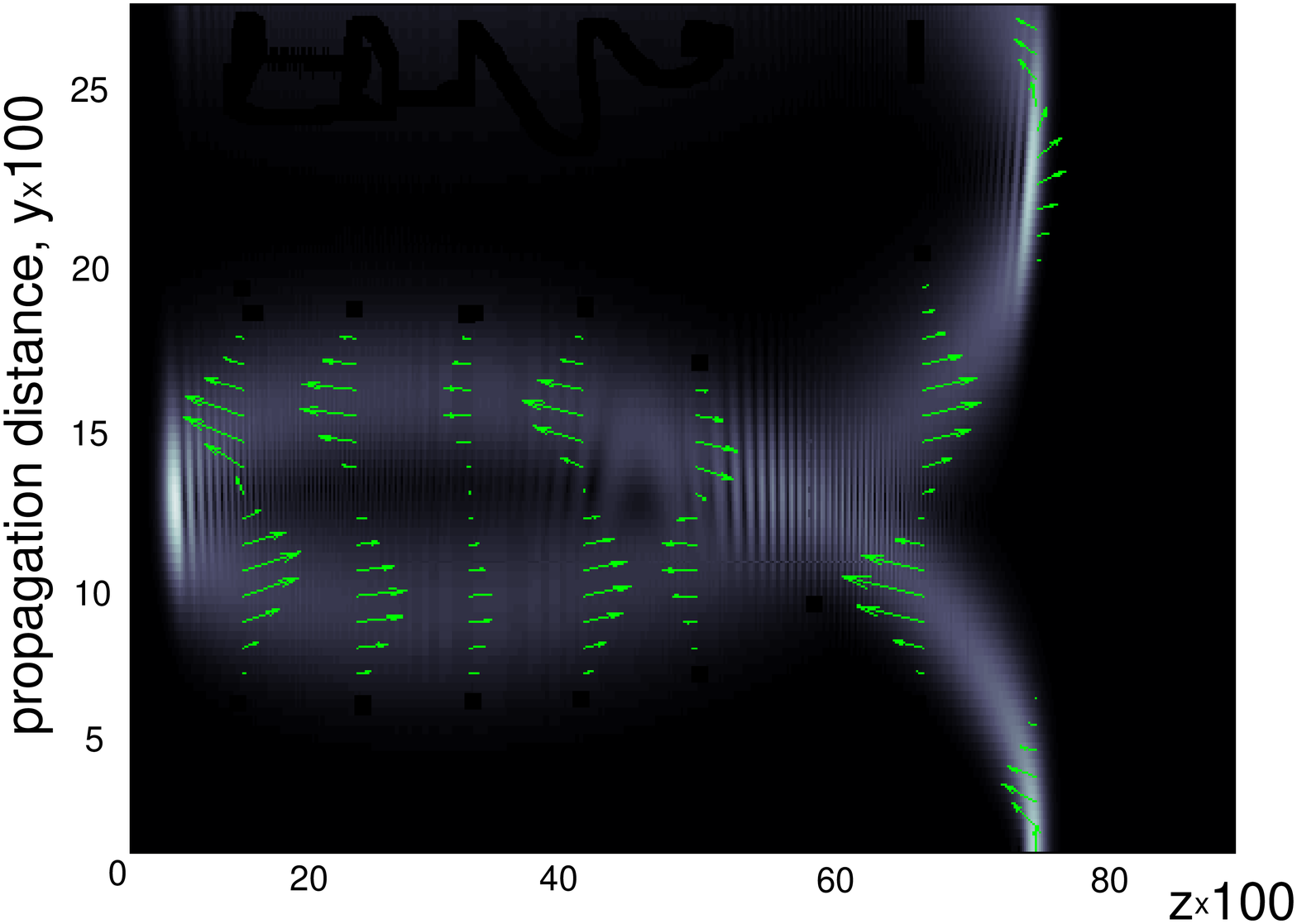}{fig_Poyting_distribution}{(Color online) Distribution of the averaged field amplitude and Poynting vector in the chirped structure. Arrows show the direction of the energy flow. }

We also calculate the average $y$-component of the Pointing vector, defined as $<S_y>$, on the unit cell and show its dynamics across the structure for the mode with the propagation constant $k_y = 1.838$, see Fig.~\rpict{fig_amplitudes}. It is clearly seen that at the left boundary of the band the energy flow direction is opposite to that near the right boundary, and it vanishes in the center of the structure. This dynamics is consistent with the behavior predicted within geometric optics approximation in Secs.~\ref{sec_structure} and \ref{sec_geometry}. We note that the total energy flow, i.e. the
integrated flow across the structure, remains positive for this mode and for all modes contributing to the beam spectrum.

\section{Conclusions}

We have presented a systematic analysis of the propagation of electromagnetic waves in chirped periodic structures composed of two kinds of alternating layers, the layers of negative-index metamaterial and the layers of conventional dielectrics, under the condition of the zero averaged refractive index. We have considered the chirp in the structure parameters introduced by varying the thickness of all layers linearly across the structure, and we have applied the methods of the geometric optics for analyzing the beam propagation and Bloch oscillations in such infinite composite structures. For the adiabatically changing period of structure, we have predicted the beam self-crossing in the bands, and have found that the energy flow in such multi-layer structures with metamaterials may have the opposite direction at the band edges, in a sharp contrast to the similar structures composed of conventional dielectrics. This novel effect of the beam curling can be useful for the beam steering in the transmission band.

\section*{Acknowledgements}

This work has been supported by the Australian Research Council through the Discovery projects.

\renewcommand{\theequation}{A\arabic{equation}}

\setcounter{equation}{0}

\section*{APPENDIX A}

In this Appendix we discuss the relation between the Bloch amplitudes $A_m$ and $B_m$ for the m-th layer and the amplitudes $A_0$ and $B_0$   in the 0-th layer.

We describe the field in the m-th unit cell of the periodic structure with the period $\Lambda_m$ as a Bloch wave $E_m(z,y)$ composed of the counter-propagating components with the amplitudes $A_m$ and $B_m$, see Eq.~(\ref{eq_eigenfield}). We consider the boundary between m-th and (m+1)-th unit cells, i.e. $z=z_{m+1}$, see Fig.~\rpict{fig_stacks}. The boundary condition at this interface requires that
\bea \label{eq_boundary_conditions}
  E_m(z_{m+1})=E_{m+1}(z_{m+1}) \; \; \; \; \; \; \; \; \; \; \; \; \; \; \; \; \\
\frac{1}{\mu_l}\left.\frac{\partial E_m(z)}{\partial
z}\right|_{z_{m+1}}=\frac{1}{\mu_r}\left.\frac{\partial
E_{m+1}(z)}{\partial z}\right|_{z_{m+1}}. \nonumber \eea

Substituting Eq.~(\ref{eq_eigenmode}) into Eq.~(\ref{eq_boundary_conditions}), we obtain:
\bea \label{eq_full_boundary_conditions}
\begin{array}{c}
  A_m(a_l^++b_l^+)+B_m(a_l^-+b_l^-)=  \\
  =A_{m+1}(a_r^+e^{ik_r\Lambda}+b_r^+e^{-ik_r\Lambda})e^{-iK_b\Lambda}+ \\
  +B_{m+1}(a_r^-e^{ik_r\Lambda}+b_r^-e^{-ik_r\Lambda})e^{iK_b\Lambda},  \\ \\
  \frac{k_l}{\mu_l}\left[A_m(a_l^+-b_l^+)+B_m(a_l^--b_l^-)\right]= \\
  =\frac{k_r}{\mu_r}\left[A_{m+1}(a_r^+e^{ik_r\Lambda}-b_r^+e^{-ik_r\Lambda})e^{-iK_b\Lambda}\right.+ \\
  +\left.B_{m+1}(a_r^-e^{ik_r\Lambda}-b_r^-e^{-ik_r\Lambda})e^{iK_b\Lambda} \right].
\end{array}
\eea
where $\pm$ corresponds to the forward and backward propagating Bloch waves, respectively.
Equation~(\ref{eq_full_boundary_conditions}) can be presented in the matrix form,
\bea
D_{ml}\left(
\begin{array}{c}
A_m\\
B_m
\end{array}
\right) = D_{mr} \left(
\begin{array}{c}
A_{m+1}\\
B_{m+1}
\end{array}
\right).
\eea
Using this matrix equation, we can present the field amplitudes at the
m-th layer thought the corresponding amplitudes at the 0-th layer,
\bea
\left(
\begin{array}{c}
A_0 \\
B_0
\end{array}
\right) = \left[\prod \limits_{i=1}^m D^{-1}_{il}D_{ir}\right]
\left(
\begin{array}{c}
A_m \\
B_m
\end{array}
\right).
\eea

\end{sloppy}

\begin{thebibliography} {10}

\bibitem{Veselago:1967-2854:SPSS} V.~G. Veselago,
Sov. Phys. Solid State {\bf 8}, 2854 (1967).

\bibitem{Shelby} R.A. Shelby, D.R. Smith, and S. Shultz, Science {\bf 292}, 77 (2001)

\bibitem{Smith:2004-788:SCI} D. R. Smith, J. B. Pendry, and M.C.K. Wiltshire,
Science {\bf 305}, 788 (2004).

\bibitem{Pendry:2000-3966:PRL} J. B. Pendry,
Phys. Rev. Lett. {\bf 85}, 3966 (2000).

\bibitem{cloak_1} J.B. Pendry, D. Schurig, and D.R. Smith,
Science {\bf 312}, 1780 (2006).

\bibitem{cloak_2} D. Schurig, J.J. Mock, B.J. Justice, S.A. Cummer, J.B. Pendry,
A.F. Starr, and D.R. Smith,
Science {\bf 314}, 977 (2006).

\bibitem{cloak_3} U. Leonhardt,
Science {\bf 312}, 1777 (2006).

\bibitem{Shadrivov:2003-3820:APL} I.V. Shadrivov, A.A. Sukhorukov, and Yu.S. Kivshar,
Appl. Phys. Lett. {\bf 82}, 3820 (2003).

\bibitem{Shadrivov:2005-193903:PRL} I.V. Shadrivov, A.A. Sukhorukov, Yu.S. Kivshar,
Phys. Rev. Lett. {\bf 95}, 193903 (2005).

\bibitem{Wu:2003-235103:PRB} L. Wu, S. He, L. Shen,
Phys. Rev. B {\bf 67}, 235103 (2003).

\bibitem{Li:2003-083901:PRL} J. Li, L. Zhou, C.T. Chan and P. Sheng,
Phys. Rev. Lett. {\bf 90}, 083901 (2003).

\bibitem{anderson} A.A. Asatryan, L.C. Botten, M.A. Byrne, V.D. Freilikher, S.A. Gredeskul,
I.V. Shadrivov, R.C. McPhedran, and Yu.S. Kivshar,
Phys. Rev. Lett. {\bf 99}, 193902 (2007).

\bibitem{Bloch_oscill} C. M. de Sterke, J. N. Bright, P. A. Krug, and T. E. Hammon,
Phys. Rev. E {\bf 57}, 2365 (1998).

\bibitem{Wilkinson:2002-056616:PRE} P. B. Wilkinson,
Phys. Rev. E {\bf 65}, 056616 (2002).

\bibitem{our_oe} A.R. Davoyan, I.V. Shadrivov, A.A. Sukhorukov, and Yu.S. Kivshar,
Opt. Express {\bf 16}, 3299 (2008).

\bibitem{Yeh:1979-742:JOSA} P. Yeh,
J. Opt. Soc. Am. {\bf 69}, 742 (1979).

\bibitem{Yeh:1988:Optical_Waves} P. Yeh, {\em Optical Waves in Layered Media} (John Wiley \& Sons, New York,
1988).

\bibitem{Born:1964:Principles_of_optics} M. Born and E. Wolf, {\em Principles of Optics: Electromagnetic
Theory of Propagation, Interference and Diffraction of Light} (Cambridge University Press, UK, 2002).


\bibitem{Landau:2001:Theoretical_Physics} L. D. Landau and E M Lifshitz {\em {Theoretical Physics}}
   (Fyz. Mat. Lit, 2001) (in Russian) vol. {\bf 1} and {\bf 2}.

\bibitem{Kravcov:1980:Geometric_optics_nhm_media} Yu. A. Kravtzov and Yu. I. Orlov,
   {\em Geometric Optics of Inhomogeneous Media} (Springer-Verlag, Berlin, 1990).

 \bibitem{Ko:1988-1863:JOSA} D. Y. K. Ko and J. R. Sambles,
 J. Opt. Soc. Am. {\bf 5}, 1863 (1988).


\end{thebibliography}
\end{document}